\documentclass[aps,pra,twocolumn,letterpaper,twoside,nobalancelastpage,superscriptaddress,amsmath,amssymb,floatfix,citeautoscript,showpacs,footinbib]{revtex4-1}
\usepackage{graphicx}
\usepackage{bm} %bold math symbols
\usepackage{times}

\usepackage{color}
\definecolor{darkblue}{rgb}{0.,0.,0.6}
\usepackage{framed}
\usepackage[
breaklinks,
colorlinks=true,
linkcolor=darkblue,
citecolor=darkblue,
urlcolor=darkblue]{hyperref}

\DeclareMathOperator{\tr}{tr}
\DeclareMathOperator{\erf}{Erf}

\begin{document}

\newcommand {\be}{\begin{equation*}}
\newcommand {\ee}{\end{equation*}}
\newcommand {\bee}{\begin{equation}}
\newcommand {\eee}{\end{equation}}

\def \Ns {\mathbb{N}}
\def \Rs {\mathbb{R}}
\def \Zs {\mathbb{Z}}
\def \Qs {\mathbb{Q}}
\def \Cs {\mathbb{C}}
\def \id {\mathbb{I}}

\def \bfq {{\bf q}}
\def \bfp {{\bf p}}
\def \bfx {{\bf x}}
\def \bfy {{\bf y}}
\def \bfz {{\bf z}}
\def \bfr {{\bf r}}
\def \bfk {{\bf k}}
\def \bfn {{\bf n}}
\def \bfb {{\bf b}}

\def \bfA {{\bf A}}
\def \bfX {{\bf X}}
\def \bfR {{\bf R}}
\def \bfP {{\bf P}}
\def \bfT {{\bf T}}
\def \bfsigma {\boldsymbol{\sigma}}
\def \bfpi {\boldsymbol{\pi}}

\def \calA {\mathcal{A}}
\def \calB {\mathcal{B}}
\def \calD {\mathcal{D}}
\def \calF {\mathcal{F}}
\def \calH {\mathcal{H}}
\def \calI {\mathcal{I}}
\def \calN {\mathcal{N}}
\def \calO {\mathcal{O}}
\def \calR {\mathcal{R}}
\def \calS {\mathcal{S}}
\def \calV {\mathcal{V}}
\def \calW {\mathcal{W}}
\def \calK {\mathcal{K}}

\def \olu {\overline{u}}
\def \olw {\overline{w}}
\def \olz {\overline{z}}
\def \olpartial {\overline{\partial}}
\def \oli {\overline{\imath}}
\def \olj {\overline{\jmath}}
\def \olk {\overline{\k}}

\def \hatq {\widehat{q}}
\def \hatp {\widehat{p}}
\def \hata {\widehat{a}}
\def \hatadag {\widehat{a}^{\dagger}}
\def \wtN {\widetilde{N}}

\def \ve {\varepsilon}
\def \vth {\vartheta}

\title{Quantum geometry and stability of the fractional quantum Hall effect in the Hofstadter model}

\author{T. S. Jackson}
\email[]{tsjackson@physics.ucla.edu}
\affiliation{Department of Physics and Astronomy, University of California at Los Angeles, 475 Portola Plaza, Los Angeles, California 90095, USA}

\author{David Bauer}
\affiliation{Department of Physics and Astronomy, University of California at Los Angeles, 475 Portola Plaza, Los Angeles, California 90095, USA}

\author{Rahul Roy}
\affiliation{Department of Physics and Astronomy, University of California at Los Angeles, 475 Portola Plaza, Los Angeles, California 90095, USA}

\date{\today}
\begin{abstract}

We study how the stability of the fractional quantum Hall effect (FQHE) is influenced by the geometry of band structure in lattice Chern insulators. We consider the Hofstadter model, which converges to continuum Landau levels in the limit of small flux per plaquette. This gives us a degree of analytic control not possible in generic lattice models, and we are able to obtain analytic expressions for the relevant geometric criteria. These may be differentiated by whether they converge exponentially or polynomially to the continuum limit. We demonstrate that the latter criteria have a dominant effect on the physics of interacting particles in Hofstadter bands in this low flux density regime. In particular, we show that the many-body gap depends monotonically on a band-geometric criterion related to the trace of the Fubini-Study metric.
\end{abstract}


\maketitle

\section{Introduction}

\subsection{Overview}

% why FCIs & what needs to be done
The theoretical proposal and numerical observation of time-reversal breaking lattice models exhibiting fractional quantum Hall (FQH) phases, reviewed in Refs.~\onlinecite{Parameswaran:2013uf,Bergholtz:2013ue}, have sparked intense interest in these models (``fractional Chern insulators'' or FCIs) as a promising venue for observing FQH physics without the need for large external magnetic fields. The majority of our theoretical understanding of the the fractional quantum Hall effect (FQHE) has been framed in the context of continuum Landau levels, which occupy a special and highly atypical point in the space of all single-particle Hamiltonians. It is a matter of both theoretical and experimental interest to identify which of the special properties of Landau levels (LLs) are essential to the existence of the FQHE: as FCIs rapidly move towards experimental reality, one would like to easily identify experimental parameters where FQH-like phases are most robust.

%``some progress towards understanding FCIs was obtained in R, J in ''

% why band geometry

% Don't mention GMP here. Remove analogy. 
% restore ``Relevant to these aims, .. from David intro. ''
% More broadly, one would like to understand the wide literature on different FCI models in terms of a unifying theoretical framework,

Relevant to these aims are several recent studies investigating the role of the algebra of band-projected density operators in an FCI in shaping the excitation spectrum of FQH phases~\cite{Murthy:2011vi,*Murthy:2012ta,*Murthy:2014fg,Goerbig:2012cz,Parameswaran:2012cu,Dobardzic:2013gc,*Repellin:2014fn,Roy:2012vo}. Intriguingly, these results point to a role for band geometry, i.e., \textit{non}-topological properties of Chern bands, in determining the stability of FQH states in FCIs; in particular, the Fubini-Study metric describing the embedding of the occupied bands in the Hilbert space of all bands appears naturally when one considers a long-wavelength expansion of this operator algebra \cite{Roy:2012vo}. That reference gave qualitative arguments linking properties of this algebra with the stability of FQH-like states which were numerically investigated in \cite{Jackson:tsinVXaa}. The results obtained there suffered from the limitation that underlying band geometry could only be varied indirectly; moreover, one would like to be able to unify these numerical examples with a theoretical picture capable of making analytic statements.

%> also highliht here what prev work did *NOT* do. Mention previous papers are ROy, Jackson, but also limitations. Want nalytic statements, better control over band geo. Point: a theoretical picture is missing. Claims should be equivalent to those made in cover.

%Why Hofstadter
%In the present work, we investigate conditions proposed in \cite{Roy:2012vo} for the existence of a generalized GMP algebra in the context of the Hofstadter model \cite{Hofstadter:1976js}. 

For these reasons, in this article we investigate the conditions proposed in \cite{Roy:2012vo} in the context of the Hofstadter model \cite{Hofstadter:1976js}. This provides an ideal laboratory for studying the interplay between band geometry and FQH physics, due to the fact that it offers a controlled limit (that of vanishing flux per plaquette) which converges to continuum LLs, which offer optimal conditions for FQH states (in a sense to be made precise below) and allows us to obtain analytic results not possible for generic FCIs. We note that, although the Hofstadter model involves a net magnetic flux, numerical studies \cite{Scaffidi:2012dx,Wu:2012ky} have shown its FQH states may be adiabatically connected to those in FCIs with no net flux. A significant literature on the FQHE in the Hofstadter model exists \cite{Sorensen:2005bt,Palmer:2006km,Hafezi:2007gz,Moller:2009ir,Kapit:2010ky,Sterdyniak:2012jo,Scaffidi:2014tg,Harper:2014vi}. The Hofstadter model is also a subject of ongoing experimental work: Hostadter bands have recently been realized with cold atoms in optical lattices \cite{Aidelsburger:2013ew,Miyake:2013jw,Aidelsburger:2014hm} and in graphene superlattices \cite{Dean:2013bv,Ponomarenko:2013hl}.

In what follows, we show that the band-geometric criteria proposed in \cite{Roy:2012vo} meaningfully quantify the stability of FQH phases as one perturbs the band geometry of the Hofstadter model away from the continuum limit. We show that momentum-space fluctuations of the Berry curvature and Fubini-Study metric have exponential convergence to their LL limit, while a third, nontrivial condition related to the averaged trace of the Fubini-Study metric has slower, polynomial convergence. Numerical exact diagonalization studies of several different FQH states show that the many-body gap depends monotonically on this third criterion, even when momentum-space fluctuations are negligible. This constitutes strong evidence that the band geometry plays a key role in the stability of the FQHE in settings where lattice effects are important.

\subsection{Geometry of Chern bands}

In this section and the next, we present the background material necessary to formulate our hypothesis for the stability of FQH-like states in the Hofstadter model and fix notational conventions which will be used in subsequent calculations. 

In momentum space, we write the band Hamiltonian of a lattice model as $H(\bfk) = \sum_{\alpha =1}^{N} E_{\alpha}(\bfk) | \bfk, \alpha \rangle \langle \bfk, \alpha |$, where $\alpha = 1, \ldots, N$ indexes the bands. A band is parameterized by its Bloch function $u^\alpha(\bfk)$, an $N$-component vector in terms of which $ | \bfk, \alpha \rangle = \sum_{b=1}^{N}  u^\alpha_b(\bfk) |\bfk, b \rangle$, where $|\bfk, b \rangle$ is the Fourier transform of the $b$th basis orbital. Topological order in a band is indicated by nonzero values of a corresponding topological invariant, which for two-dimensional systems with broken time-reversal symmetry is the first Chern number 
\begin{equation}
c_1 = A_{BZ}\langle B_\alpha \rangle / 2 \pi,
\end{equation} 
where $A_{BZ}$ is the area of the Brillouin zone (BZ), $\langle \cdots \rangle$ denotes the average over the BZ, normalized so $\langle 1 \rangle = 1$, and the Berry curvature \cite{Simon:1983du,Berry:1984ka} of the band $\alpha$ in terms of Bloch functions is 
\begin{equation}
\label{eq_curvature_def}
B_\alpha(\bfk) = -i \sum_{b=1}^{N} \left( \frac{\partial u_b^{\alpha \ast}}{\partial k_x}\frac{\partial u_b^{\alpha}}{\partial k_y} -  \frac{\partial u_b^{\alpha \ast}}{\partial k_y}\frac{\partial u_b^{\alpha}}{\partial k_x} \right).
\end{equation}
One may additionally define a ``quantum metric'' on the BZ by introducing the Fubini-Study metric on the band Hilbert space \cite{Provost:1980hs} and using $u^\alpha_b(\bfk)$ to pull it back to the BZ, yielding
\begin{align}
\label{eq_metric_def}
g^\alpha_{ij}(\bfk) &= \frac{1}{2} \sum_{b=1}^{N} \left[ \left( \frac{\partial u_b^{\alpha \ast}}{\partial k_i}\frac{\partial u_b^{\alpha}}{\partial k_j} + \frac{\partial u_b^{\alpha \ast}}{\partial k_j}\frac{\partial u_b^{\alpha}}{\partial k_i} \right) \right. \\\
{}& \quad - \sum_{c=1}^{N} \left.\left(\frac{\partial u_b^{\alpha \ast}}{\partial k_i}u^\alpha_b u^{\alpha \ast}_c \frac{\partial u_c^{\alpha}}{\partial k_j} + \frac{\partial u_b^{\alpha \ast}}{\partial k_j}u^\alpha_b u^{\alpha \ast}_c \frac{\partial u_c^{\alpha}}{\partial k_i} \right) \right]. \nonumber
\end{align}
We emphasize that, despite their mathematical origin, the curvature and metric are quite natural physical quantities: observable effects of Berry curvature are reviewed in \cite{Xiao:2010kw}, and the quantum metric is experimentally accessible through current noise measurements \cite{Neupert:2013tp}. 

In the present work, we consider three geometric conditions for the stability of FCIs involving $B_{\alpha}$ and $g^{\alpha}$ which were first proposed in Ref.~\onlinecite{Roy:2012vo} (in a refinement of previous arguments \cite{Parameswaran:2012cu,Parameswaran:2013uf}). These are that (i) the Berry curvature \eqref{eq_curvature_def} is constant over the BZ, (ii) all components of the quantum metric \eqref{eq_metric_def} are constant over the BZ, and (iii) the following quantity 
\begin{equation}
\label{eq-detineq}
D(\bfk) \equiv \det g^\alpha(\bfk) - \frac{B_\alpha(\bfk)^2}{4} \geq 0
\end{equation}
vanishes everywhere, i.e. the metric determinant inequality \eqref{eq-detineq} is saturated. In Ref.~\onlinecite{Roy:2012vo} a similar inequality involving the trace of the metric
\begin{equation} 
\label{eq-trineq}
T(\bfk) \equiv \tr g^\alpha(\bfk) - |B_\alpha(\bfk)| \geq 0
\end{equation}
was proved, and it was shown that $T(\bfk) = 0$ automatically implies $D(\bfk)=0$. 

Conditions (i), (ii) and (iii) were not chosen arbitrarily: the main result of Ref.~\onlinecite{Roy:2012vo} is that they are sufficient to guarantee that the set of band-projected density operators in an FCI is closed under commutation, and obeys a generalization of the Girvin-MacDonald-Platzman algebra \cite{Girvin:1986bu} arising in the continuum FQHE. Density-density interactions are essential to the stability of any FQH phase, so the more accurately the algebra is reproduced (specifically, as measured by the conditions), the more stable FQH-like phases should be: we refer to this as the ``band geometry hypothesis.'' 

The validity of the band geometry hypothesis has been tested for a variety of FQH-like phases appearing in several different FCI models in \cite{Jackson:tsinVXaa}. We take a different approach in the present work: we consider a sequence of lattice models (the Hofstadter model at $\phi = 1/N$, defined below) which converges to continuum FQHE physics as $N \to \infty$, which lets us \textit{quantitatively} investigate how the stability of FQH states depends on band geometry in a way that was not possible in Ref.~\onlinecite{Jackson:tsinVXaa}.

\subsection{Hofstadter model in the limit of small flux per plaquette}

The quantum dynamics of an electron on a two-dimensional square lattice in the presence of a transverse magnetic field was first studied by Harper \cite{Harper:1955ft}, and its self-similar band structure was further elucidated by Azbel \cite{Azbel:1964tk}. With these caveats, for brevity we refer to the following tight-binding Hamiltonian as the Hofstadter model \cite{Hofstadter:1976js}:
\begin{equation}
\label{eq-hof}
H = -t \sum_{\langle i j \rangle} \left[ c_{i}^{\dagger} c_{j} \exp \left( 2 \pi i \int_{\bfr_{i}}^{\bfr_{j}} \! \bfA \cdot d\ell \right) + \text{h.c.} \right],
\end{equation}
with $i,j$ indexing sites on a square lattice with lattice constant $a$. In the Landau gauge, $\mathbf{A} = Bx\,\mathbf{\hat{y}}$ and all eigenstates have trivial $y$ dependence $\psi(x= na, y) = e^{i k_y y}\psi_{n}$, with the remaining $x$ dependence obeying the discrete Harper equation
\begin{equation}
\label{eq-harper}
\psi_{n+1} + \psi_{n-1} +  2\cos(2 \pi \phi n - k_{y}) \psi_{n} = \varepsilon \psi_{n},
\end{equation}
where $\varepsilon$ is the energy in units of $t$ and $\phi$ is the magnetic flux per lattice cell in units of the flux quantum $\phi_{0}= h/e$. For  rational $\phi = P/Q$, \eqref{eq-harper} is invariant under $n \to n+Q$ and one may apply Bloch's theorem over the enlarged magnetic unit cell consisting of $Q \times 1$ lattice plaquettes.

In \cite{Harper:2014vi}, the Hofstadter model was studied perturbatively for $\phi$ near rational values. We recall some results relevant to the case of interest here, $\phi = 1/N$: the $N \to \infty$ limit of the Hofstadter model is the continuum Landau level (LL) problem in a precise sense. We first recall that in the Landau gauge, states in the LL problem are spanned by two commuting free boson algebras generated by $a, a^{\dagger}, b, b^{\dagger}$, where the first pair act as ladder operators for the LLs and the second pair raise and lower momentum within a LL. A major result of \cite{Harper:2014vi} is that Hofstadter bands may be written as superpositions of LLs by a unitary change of basis, given for large $N$ by
\begin{widetext}
\begin{align}
\label{eq-U3}
U^{\dagger} &= \exp\left[\left(\frac{1}{96}\frac{\pi}{N} + \frac{1}{128}\frac{\pi^2}{N^2} + \frac{37}{6912}\frac{\pi^3}{N^3}\right)\left( a^{\dagger 4} - a^4\right) +
%\right. \nonumber \\ {} & \qquad \qquad + \left.
\left(\frac{1}{320}\frac{\pi^2}{N^2} + \frac{7}{1152}\frac{\pi^3}{N^3}\right)\left(a^{\dagger 5} a - a^{\dagger}a^5\right) \right. \nonumber \\
{} & \qquad \qquad + \left. \frac{7}{6912}\frac{\pi^3}{N^3}\left(a^{\dagger 6} a^2 - a^{\dagger 2}a^6\right)- \frac{11}{215040}\frac{\pi^3}{N^3}\left( a^{\dagger 8} - a^8\right) +  O\left(N^{-4} \right)\right]. 
\end{align}
\end{widetext}

% The lattice potential breaks the full rotational symmetry of the continuum problem, but angular momentum non-conserving terms are exponentially small in $N$, as are gauge-dependent terms (see Sec.~VI of \cite{Harper:2014vi}).

\subsection{WKB Hofstadter wavefunctions}
\label{sec-wkb-prev}

In this section we recall details of the Hofstadter eigenfunctions in the WKB approximation \cite{Watson:1991gy,Harper:2014vi,harpernotes} (not all of which have been published) which are needed for our calculation in Sec.~\ref{sec-wkb}. In the large-$N$ limit, we may regard the lattice site index $n$ in the Harper equation \eqref{eq-harper} as a continuous variable $x = n/N$. Replacing discrete differences by derivatives, we obtain a second-order differential equation in  which permits a WKB approximation:
\begin{equation}
\label{eq-harper-diffeq}
\frac{1}{N^2}\psi''(x) + \left[(\varepsilon + 2) + 2\cos(2\pi x)\right]\psi(x) = 0.
\end{equation}
Here we have set $k_{y} =0$, since increasing $k_{y}$ is equivalent to a translation in $x$ in the Landau gauge.

The classical turning points of \eqref{eq-harper-diffeq} occur at $x=m \pm \beta$ for integer $m$ and $\beta = (1/2\pi)\cos^{-1}\left(-\varepsilon/2 - 1\right)$; in the large-$N$ limit $\beta \sim 1/\sqrt{2\pi N}$, so that in the large-$N$ limit the two first-order turning points at $m-\beta$, $m+\beta$ coincide at the minimum of the potential to form a single second-order turning point (so that the region corresponding to the oscillatory WKB solution vanishes). We use a quadratic approximation to the cosine potential near the turning points and use the WKB solution only in the classically forbidden region.

We define the Bloch wavefunction piecewise in three regions of the magnetic unit cell, with
\begin{flalign*}
u_\mathbf{k}(x) = \left\{
     \begin{array}{lr}
      \multicolumn{2}{l}{A U(a,\xi) + B V(a,\xi),}\\
      \multicolumn{2}{r}{0 \leq x < \alpha \beta;}\\
      \multicolumn{2}{l}{C\psi^{+}_{\text{exp}}(x) + D\psi^{-}_{\text{exp}}(x),}\\
      \multicolumn{2}{r}{\alpha \beta \leq x < 1 - \alpha \beta;}\\
      \multicolumn{2}{l}{F U(a,\xi-\xi(1)) + G V(a,\xi-\xi(1)),\phantom{XXX}}\\
      \multicolumn{2}{r}{1 - \alpha \beta \leq x < 1.}
     \end{array}
   \right.
\end{flalign*}
The first and third regions are in the vicinity of the second order turning point (integer $x$), with the solution given by parabolic cylinder functions  $U(a,\xi)$, $V(a,\xi)$ as discussed in Ref.~\onlinecite{Harper:2014vi}, where $a =\frac{N}{4\pi}\left(\varepsilon + 4\right)$ and $\xi(x) = 2\sqrt{\pi N}x$ are suitably rescaled variables. The second region is the classically forbidden region, where the solution decays exponentially. There is no canonical point at which the wavefunctions from different regions should be matched, which is expressed by a dimensionless free parameter $\alpha >1$. Its value does not change the qualitative nature of our conclusions, and we use the value $\alpha = 2.3$ adopted in Ref.~\onlinecite{Harper:2014vi} so that our results are compatible with that work.

%To lowest order, this reproduces the Schr{\"o}dinger equation for Landau levels in the Landau gauge, with lattice effects manifesting at higher orders in $N$. 

%\section{WKB metric oscillations}
%In this section we give details on the calculation of the asymptotic $N$ dependence of the amplitude of BZ fluctuations of metric components, using the WKB approximation to the Bloch wavefunction of the lowest Hofstadter band \cite{Watson:1991gy,Harper:2014vi,harpernotes}.

%As explained in the main text, in the large-$N$ limit we approximate the discrete Harper equation by a second-order differential equation, which we then solve with a WKB expansion in the small parameter $1/N$. 

%Expanding the cosine around $x=0$ and introducing rescaled variables  yields
%\begin{equation}
%\label{eq-parabolic-cyl}
%\psi''(\xi) - \left(\frac{1}{4}\xi^2 + a\right)\psi(\xi) = 0,
%\end{equation}
%the solutions of which are parabolic cylinder functions The limit $N \to \infty$ corresponds to $a\rightarrow -1/2$, due to the fact that $\varepsilon \sim -4 + 2\pi/N + \ldots$; in this limit, \eqref{eq-parabolic-cyl} reduces to the Schr{\"o}dinger equation for a quantum harmonic oscillator in its ground state, i.e. the lowest Landau level.

% The exponentially decaying part of the WKB solution is 
%\begin{equation}
%\psi^{\pm}_{\text{exp}}(x) = \frac{1}{\sqrt{\sinh \tilde{p}(x)}} \exp\left[\pm N \int^x_{\beta_{L,R}} \tilde{p}(y) \, dy\right],
%\end{equation}
%where $\beta_{L,R}$ are turning points on either side of a unit cell. In general, 

The Bloch function in the second, classically forbidden region is \cite{harpernotes}
\begin{align}
\label{eq-bloch2}
u_{\bfk, \text{II}}(x) = \frac{Ae^{-\tilde{\sigma}N/2}}{\sqrt{\sinh \tilde{p}(x)}} &\left[ e^{-ik_{x}xN} u_{-}(x-k_{y}/2\pi) \right.  \\ & \quad \left. + e^{ik_{x}(1-x)N} u_{+}(x-k_{y}/2\pi) \right],\nonumber
\end{align}
(correcting a misprint in Ref.~\onlinecite{Harper:2014vi}), where $u_{\pm}(x)$ is
\begin{equation}
\label{eq-upm-def}
u_{\pm}(x) =  \exp\left[\pm N \int^x_{1/2} \tilde{p}(y)\, dy\right],
 \end{equation}
and for future convenience we have defined 
\begin{equation}
\tilde{p}(x)  = \cosh^{-1}\left[-\varepsilon/2- \cos(2 \pi x) \right].
\end{equation} 
We have the symmetries $u_{-}(x) = u_{+}(1-x)$ and $u_{-}(x) = 1/u_{+}(x)$. Effects which are nonperturbative in $N$ in \eqref{eq-bloch2} and what follows are governed by powers of $e^{-\tilde{\sigma}N}$, where we define
\begin{equation}
\label{eq-ts-def}
\tilde{\sigma} = \int_{\beta}^{1-\beta} \! \tilde{p}(x) dx \approx 1.166 - \frac{0.208}{\sqrt{N}} - \frac{2.227}{N} + \ldots,
\end{equation}
with the asymptotic value given by $4/\pi$ times Catalan's constant, $G \equiv \sum_{n=0}^{\infty}(-1)^{n}/(2n+1)^{2}$. In what follows we only keep contributions of $O(e^{-\tilde{\sigma}N})$.

Finally, in \eqref{eq-bloch2} $A$ is the normalization of the entire Bloch function $u_{\bfk}$. This is dominated by contributions from the first and third regions, where $|u_{\bfk}|^{2}$ has its maxima \cite{harpernotes}:
\begin{equation}
\label{eq-a-def}
|A|^{2}\approx \frac{1}{N} \sqrt{\frac{\pi}{e}} \frac{1}{\erf(\alpha)}.
\end{equation}
Contributions from region II enter $|A|^{2}$ at $O(N^{-2})$, but we've found that the coefficient is numerically negligible at $\alpha = 2.3$, and we drop it in the rest of the analysis to avoid the complications involved in attempting to satisfy the self-consistent set of equations that would result.

Note that in \eqref{eq-bloch2} the $k_{y}$ dependence has been re-inserted ``by hand'', under the assumption of continuous translational dependence. As a consequence, we cannot recover any information about fluctuations in the $k_{y}$ direction and must appeal to symmetry arguments presented in section~\ref{app-symmetry} to constrain the full functional dependence from the $k_{x}$ behavior. 

%[mention LL mixing U, and how it's approximately gauge-independent]

\subsection{Summary}
Our results are grouped into two sections. In Section~\ref{sec-singleparticle}, we find analytical expressions for aspects of the quantum metric of the lowest Hofstadter band in the limit of small flux per plaquette $\phi$. Considering lattice effects as perturbations of the lowest Landau level in this limit, we calculate the average of the metric over the Brillouin zone and the $\bfk$ dependence of fluctuations about this average. We also provide nonperturbative results, in particular statements about the $\bfk$ dependence of the metric from symmetry considerations. The results we obtain in Section~\ref{sec-singleparticle} together with similar results for the Berry curvature found in Ref.~\onlinecite{Harper:2014vi} allow us to describe the behavior of the determinant and trace inequalities.

In Section~\ref{sec-multiparticle}, we study multiple FQH-like states by numerically diagonalizing many-particle interactions projected to the lowest Hofstadter band and finding the corresponding many-body spectra. Our main findings are that geometric quantities can be distinguished by their functional dependence on $N$ in the continuum limit, vanishing either exponentially or polynomially, and that the latter of these appear to have the dominant effect on the stability of fractional phases as measured by the size of the many-body gap. Our conclusions are given in Sec.~\ref{sec-discussion}. In Appendix~\ref{app-n3}, we present explicit calculations for $N=3$ which illustrate the results of Sec.~\ref{sec-singleparticle}.

%In the remainder of the paper, we investigate the band geometry of the Hofstadter model, viewed as a lattice approximation to continuum Landau levels, and the influence it has on the stability of the fractional quantum Hall effect. Section~\ref{sec-singleparticle} deals with computing the quantum metric from known single-particle wavefunctions. We first prove needed results on the symmetries of the quantum metric in momentum space in section \ref{app-symmetry}. The amplitudes of fluctuations of metric components over the BZ are computed from the WKB wavefunction in sec.~\ref{sec-wkb}, and the BZ averages of these quantities are obtained in sec.~\ref{sec-bzaverages}. Section~\ref{sec-multiparticle} contains our results on the FQHE in the Hofstadter model: we establish an important scaling relation in sec.~\ref{sec-scaling} and present results of simulations of the bosonic and fermionic Laughlin states, as well as the Moore-Read state, in sec.~\ref{sec-manybody}. Our conclusions are given in sec.~\ref{sec-discussion}.

\section{Quantum geometry of the Hofstadter model}
\label{sec-singleparticle}

\subsection{Symmetries of the quantum metric}
\label{app-symmetry}

In this section we prove results [Eqs.~\eqref{eq-xform3} and \eqref{eq-xform4}] concerning the symmetry of the quantum metric of the Hofstadter model over the Brillouin zone. This lemma is needed in the following section, where we use symmetry properties to infer the full $\bfk$ dependence of the quantum metric from the $k_{x}$ dependence of the WKB wavefunctions. The argument below closely parallels that given for the Berry curvature in Appendix~B of Ref.~\onlinecite{Harper:2014vi}, and all of the results are readily verifiable numerically. 

We proceed by comparing eigenfunctions for the Hofstadter Hamiltonian written in two different Landau gauges, $\mathbf{A}^{(Y)}=Bx \widehat{\mathbf{y}}$ and $\mathbf{A}^{(X)}=-By \widehat{\mathbf{x}}$. As indicated, in this section we identify quantities computed in the two different gauges by the superscripts $(Y)$ and $(X)$, respectively. The Harper equation in the $(Y)$ gauge is
\begin{equation}
\label{eq-harper-y}
\psi^{(Y)}_{n+1} + \psi^{(Y)}_{n-1} + \left[ 2 \cos(2\pi\phi x-k_{y}) -\varepsilon \right] \psi^{(Y)}_{n} = 0,
\end{equation}
where $\psi^{(Y)}(na,y) = e^{i k_{y}y}\psi^{(Y)}_{n}$. In the $(X)$ gauge, the Harper equation takes the form
\begin{equation}
\label{eq-harper-x}
\psi^{(X)}_{n'+1} + \psi^{(X)}_{n'-1} + \left[ 2 \cos(2\pi\phi y+k_{x}) -\varepsilon \right] \psi^{(X)}_{n'} = 0,
\end{equation}
where instead the $x$ dependence is trivial: $\psi^{(X)}(x,n' a) = e^{i k_{x}x}\psi^{(X)}_{n'}$. Equation \eqref{eq-harper-y} is transformed to \eqref{eq-harper-x} by the clockwise rotation $(x,y) \to (-y,x)$, $(k_{x},k_{y}) \to (-k_{y},k_{x})$, so this also sends eigenfunctions to eigenfunctions. Since the corresponding Bloch functions also transform trivially, this means that the behavior of the curvature and metric is determined entirely by the transformation properties of the momentum derivatives involved in their definition:
\begin{align}
\label{eq-g-refl1}
g_{xx}^{(Y)} (k_{x},k_{y}) &= g_{yy}^{(X)} (-k_{y},k_{x}), \\
g_{xy}^{(Y)} (k_{x},k_{y}) &= -g_{xy}^{(X)} (-k_{y},k_{x}), \\
g_{yy}^{(Y)} (k_{x},k_{y}) &= g_{xx}^{(X)} (-k_{y},k_{x}), \\
\label{eq-g-refl4}
B^{(Y)} (k_{x},k_{y}) &= B^{(X)} (-k_{y},k_{x}).
\end{align}

We observe that if $\psi^{(Y)}_{n}$ is a solution of \eqref{eq-harper-y} with $k = k_{y}$, then $\psi^{(Y)}_{n-1}$ (with the index $n$ interpreted cyclically modulo $N$) is also a solution with the same value of $\varepsilon$ and $k = k_{y}+ 2\pi\phi$.  Taken together with the periodicity imposed by the magnetic unit cell, this implies that $B^{(Y)}$ and each component of $g^{(Y)}$ has periodicity $2\pi/N$ in both $k_{x}$ and $k_{y}$, for any $\phi$ of the form $M/N$. A parallel argument holds for the $(X)$ gauge.

In order to compare the gauges $(Y)$ and $(X)$, we consider a real-space unit cell of $N \times N$ sites, so that each eigenstate in both gauges appears with an artificial $N$-fold degeneracy. Ordinary lattice translations by $N$ sites in the $x$ and $y$ directions commute with each other and either Hamiltonian, so we may define $V^{(Y)}_{\alpha}(\bfk)$ as the $N$-dimensional subspace spanned by eigenstates corresponding to the band $\alpha$, with momentum eigenvalues $\bfk$ now defined in the reduced BZ of size $2\pi/N \times 2\pi/N$. We may obtain a basis for $V^{(Y)}_{\alpha}(\bfk)$ for fixed $\bfk$ by taking the eigenvectors $\psi^{(Y)}(k_{x}+2\pi m /N,k_{y})$ for $m = 0, 1, \ldots, N-1$. 

The curvature and components of the metric for the subspace $V^{(Y)}_{\alpha}(\bfk)$ may be defined as the sum of the corresponding quantity over all elements of an orthonormal basis for that space, and it readily follows from the definitions \eqref{eq_curvature_def}, \eqref{eq_metric_def} that this is independent of the choice of basis (in the literature, this is sometimes referred to as a ``gauge invariance'' of the Berry curvature and quantum metric, which should not be confused with the electromagnetic gauge symmetry under discussion here.) Using the Bloch functions corresponding to the basis vectors $\psi^{(Y)}(k_{x}+2\pi m /q,k_{y})$, we have 
\begin{align}
\label{eq-xform1}
g_{ij}[V^{(Y)}_{\alpha}(\bfk)] &= Ng_{ij}^{\alpha, (Y)}(k_{x},k_{y}), \\
\label{eq-xform2}
B[V^{(Y)}_{\alpha}(\bfk)] &= NB_{\alpha}^{(Y)}(k_{x},k_{y}).
\end{align}
Corresponding statements can be made for the $(X)$ gauge. Now let $U_{YX}$ be the unitary operator mapping the Hamiltonian in the $(X)$ gauge to the Hamiltonian in the $(Y)$ gauge. This sends the subspace $V^{(X)}_{\alpha}(\bfk)$ to $V^{(Y)}_{\alpha}(\bfk)$, so we may use $U_{YX}|\psi^{(X)}(\bfk) \rangle$ as a basis for $V^{(Y)}_{\alpha}(\bfk)$ to find $g_{ij}[V^{(X)}_{\alpha}(\bfk)] = g_{ij}[V^{(Y)}_{\alpha}(\bfk)]$, and likewise for the curvature. Together with \eqref{eq-xform1}, \eqref{eq-xform2} and the $2\pi/N \times 2\pi/N$ periodicity of all quantities in question, this proves the transformation properties \eqref{eq-g-refl1}--\eqref{eq-g-refl4} apply within the gauge $(Y)$, i.e. with $(X)$ on the right-hand sides of \eqref{eq-g-refl1}--\eqref{eq-g-refl4} replaced by $(Y)$. In particular,
\begin{align}
\label{eq-xform3}
\tr g(k_{x},k_{y}) &= \tr g(-k_{y},k_{x}), \\
\label{eq-xform4}
\det g(k_{x},k_{y}) &= \det g(-k_{y},k_{x}).
\end{align}

In the following section, we will use a WKB analysis in the $(Y)$ gauge to show that the leading-order $k_{x}$ dependence of $B$, $g_{xx}$ and $g_{yy}$ is $\cos (Nk_{x})$. The symmetry \eqref{eq-xform3}, \eqref{eq-xform4} then constrains the full momentum dependence of $B$, $\tr g$ and $\det g$ to be a function of $\cos (Nk_{x}) + \cos (Nk_{y}) $ and $\cos (Nk_{x})\cos (Nk_{y})$. One may proceed to show that, of these two forms, the curvature depends \emph{only} on the sum \cite{Harper:2014vi}. Oddly, despite having the same symmetry as $B$, we have found that $\tr g$ and $\det g$ depend on both the sum and product of cosines. This may be seen from numerics, or established analytically for small $N$ using the results of Appendix~\ref{app-n3}.

\subsection{Quantum metric fluctuations}
\label{sec-wkb}

As explained above, the band geometry hypothesis predicts that the two most important factors influencing the stability of the FQHE in a lattice system are fluctuations in the Berry curvature and quantum metric, as a function of $\bfk$. The $N$ dependence of the amplitude of these fluctuations may be computed by using the WKB approximation to the Bloch wavefunction of the lowest Hofstadter band in the definitions of $B$ and $g$ [Eqs.~\eqref{eq_curvature_def} and \eqref{eq_metric_def}]. This was done for curvature fluctuations in Ref.~\onlinecite{Harper:2014vi}, but the results for fluctuations of the quantum metric we present below are new.  %We follow the notation of Appendix A of Ref.~\onlinecite{Harper:2014vi}, adding details and correcting minor errors where necessary. 

Given the Bloch function \eqref{eq-bloch2}, we can take derivatives and calculate the curvature and metric using Eqs.~\eqref{eq_curvature_def} and \eqref{eq_metric_def}, with the inner product over the band index replaced by an integral over the unit cell and $\sum_{n} \rightarrow N \int dx$. To lowest order, the parabolic cylinder function solution corresponds to continuum Landau levels, for which the metric and curvature are uniform. We therefore make the assumption %(made in \cite{Harper:2014vi} and justified by numerical evidence) 
that the leading-order oscillatory contribution to the band geometry comes solely from the WKB piece of the wavefunction in the second region, which we described in detail in Sec.~\ref{sec-wkb-prev}.

Contributions from the second term of \eqref{eq_metric_def} contain two factors of $e^{-\tilde{\sigma}N}$ and are neglected in our analysis. Differentiation of $u_{\pm}(x-k_{y}/2\pi)$ with respect to $k_{y}$ brings down factors of $N\tilde{p}(x-k_{y}/2\pi)/2\pi$, while the terms with oscillatory behavior in $k_{x}$ that we seek arise only from differentiation of the explicit exponential factors in \eqref{eq-bloch2} with respect to $k_{x}$. As a consequence, if we choose to measure the fluctuations of a quantity $X(k_{x}, k_{y})$ via $\tilde{X}(k_{x}) = X(k_{x}, 0) - X(0,0)$, all terms involving $|u_{\pm}(x)|^{2}$ cancel at $O(e^{-\tilde{\sigma}N})$. We obtain
\begin{align}
\label{eq-asymp1}
\tilde{g}_{xx} &= |A|^{2}N^{3}e^{-\tilde{\sigma}N} I_{1} \cos (Nk_{x}), \\
\tilde{g}_{xy} &= 0+ O(|A|^{4}N^{4}e^{-2\tilde{\sigma}N}), \\
\tilde{g}_{yy} &= |A|^{2}N^{3}e^{-\tilde{\sigma}N} I_{3} \cos (Nk_{x}), \\ 
\label{eq-asymp4}
\tilde{B} &=  2|A|^{2}N^{3}e^{-\tilde{\sigma}N} I_{2} \cos (Nk_{x}),
\end{align}
where $A$ is given by \eqref{eq-a-def}, $\tilde{\sigma}$ is given by \eqref{eq-ts-def} and the needed integrals are

\begin{equation}
I_{1} = 2\int^{1/2}_{\alpha\beta} \frac{x(1-x)}{\sinh \tilde{p}(x)} dx,
\end{equation}
\begin{equation}
I_{2} = 2\int^{1/2}_{\alpha\beta} \frac{\tilde{p}(x)}{(2\pi)} \frac{dx}{\sinh \tilde{p}(x)},
\end{equation}
\begin{equation}
I_{3} = 2\int^{1/2}_{\alpha\beta} \frac{\tilde{p}(x)^{2}}{(2\pi)^{2}}\frac{dx}{\sinh \tilde{p}(x)}, 
\end{equation}
where we have used the fact that all integrands are even under reflection about $x=1/2$ to fold the region of integration. The integrand obtained for $g_{xy}$ is, instead, odd about $x=1/2$ and vanishes: the only contributions to $g_{xy}$ come from terms we neglect at our level of approximation.

In Refs.~\onlinecite{Harper:2014vi,harpernotes}, $[\sinh \tilde{p}(x)]^{-1/2}$ was approximated by its average value (which is $O(1)$) and factored out of all integrations. This is a rather severe approximation, since $[\sinh \tilde{p}(x)]^{-1/2}$ diverges as $x \to \beta$, where relevant $N$ dependence enters; in fact, under the change of variables $u = \sin^{2}(\pi x)$ we find
\begin{align*}
I_{0} &= 2\int^{1/2}_{\alpha\beta} \frac{dx}{\sinh \tilde{p}(x)} \\
&\sim -\frac{1}{2\pi}\int_{\sin^{2}(\pi\alpha\beta)}^{1} \! du \frac{d}{du} \log \left[ \frac{u}{1-\sqrt{1-u^{2}}} \right],
\end{align*}
where the lower limit of integration goes to 0 as $\pi \alpha^{2}/2N$. The integrand diverges as $u \to 0$, and we have $I_{0}\sim (1/2\pi) \log N$. Note that this divergence is irrelevant in practice, as it is dominated by $e^{-\tilde{\sigma}N}$ wherever it appears.

The other integrands are progressively better behaved as $x\to \beta$, and their integrals are all finite as $N \to \infty$. Numerical integration at $\alpha = 2.3$ gives
\begin{align}
I_{1} &\approx 0.114 - \frac{0.251}{\sqrt{N}} + \frac{0.068}{N}, \\
I_{2} &\approx 0.125 - \frac{0.292}{\sqrt{N}}+ \frac{0.063}{N}, \\
I_{3} &\approx 0.0214 - \frac{0.002}{N}.
\end{align}

\begin{figure}[bht]
\centering
\includegraphics[width=3.2in]{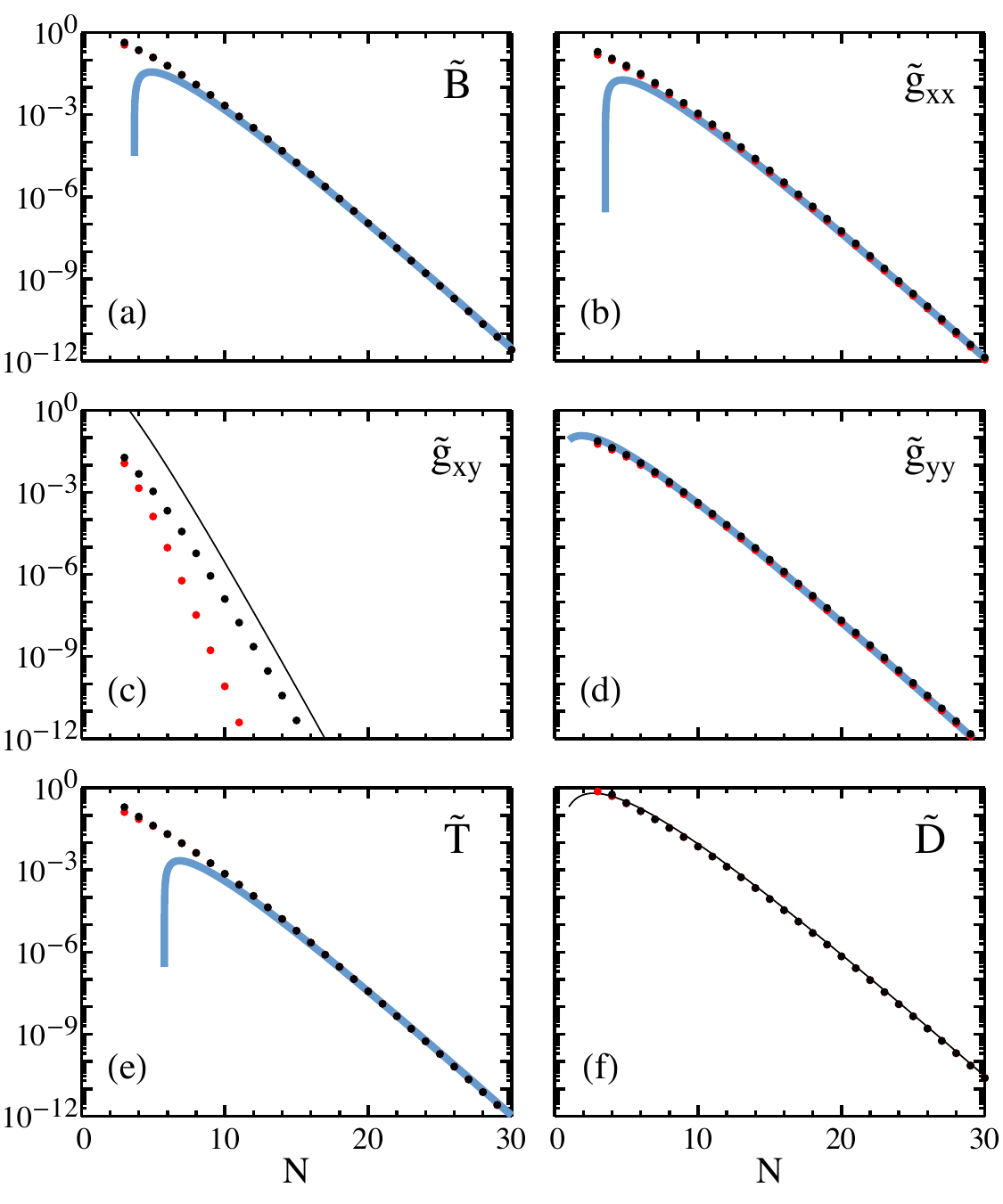}
\caption{\label{fig-fluct} (color online). Magnitude of Brillouin zone fluctuations of band-geometric quantities, as a function of $N$. Red points correspond to the amplitude of the $\cos (Nk_{x})+\cos (Nk_{y})$ term in the Fourier series, while black points correspond to the RMS fluctuation of each quantity found from numerical integration over the BZ. Blue lines show the approximation given in Eqs.~\eqref{eq-asymp1}--\eqref{eq-asymp4}. Our analysis only predicts the form of the leading $N$ dependence for $\tilde{g}_{xy}$ and $\tilde{D}$, which is shown with a thin line (the overall proportionality constant being undetermined.)}
\end{figure}

We find excellent agreement with the amplitude of the $\cos (Nk_{x})+\cos (Nk_{y})$ dependence, measured numerically, of each quantity (red points in Fig.~\ref{fig-fluct}). Unlike the Berry curvature, the metric components have additional momentum dependence which is not of this form (see Sec.~\ref{app-symmetry}), but we find that the asymptotic estimates obtained in this discussion also accurately describe the total RMS fluctuation of $g_{xx}$ and $g_{yy}$ (black points in Fig.~\ref{fig-fluct}), despite the fact that, due to the choice of Landau gauge, our analysis is unable to reconstruct the full momentum dependence of the components of the metric for large $N$. In the case of $g_{xy}$, we simply plot the leading $N$ dependence $|A|^{4}N^{4}e^{-2\tilde{\sigma}N}$ as a guide to the eye, since we cannot determine the constant of proportionality at the current level of approximation. Similar remarks hold for fluctuations in $D$; we find its $N$ dependence is well described by $(1/20)|A|^{4}N^{5}e^{-\tilde{\sigma}N}$.

%\begin{figure}[thb]
%\centering
%\includegraphics[width=3.25in]{metric-fluct2.pdf}
%\caption{\label{fig-fluct2} (color online). Magnitude of Brillouin zone fluctuations of metric-derived quantities, as a function of $N$. Red points correspond to the amplitude of the $\cos (Nk_{x})+\cos (Nk_{y})$ term in the Fourier series, while black points correspond to the RMS fluctuation of each quantity found from numerical integration over the BZ. In the first row we plot the asymptotic amplitudes obtained from eqs.~\eqref{eq-asymp1}--\eqref{eq-asymp4}, while in the second row we plot $|A|^{4}N^{5}e^{-\tilde{\sigma}N}$.}
%\end{figure}

%%%%%%%%%%%%%%%%%%%%%%%%%%%%%%%%%

\subsection{BZ-averaged metric components}
\label{sec-bzaverages}

In this section, we provide details on the perturbative calculation of BZ-averaged band-geometric quantities. The final condition appearing in the band geometry hypothesis involves the Brillouin zone average of the determinant (or trace) of the quantum metric, as opposed to the fluctuations calculated in the previous section. Since this does not require knowing the $\bfk$ dependence of these quantities, they may be computed more simply by changing to the Landau level basis using \eqref{eq-U3}, rather than by manipulating the WKB wavefunctions directly.

Using \eqref{eq-U3}, the Hofstadter ground state $| \widetilde{0} \rangle= U^{\dagger}|0\rangle$, where $|0\rangle$ is the lowest Landau level, is given in the LL basis $|m \rangle$ ($m\geq 0$) as
\begin{align}
\label{eq-perturb-gs}
| \widetilde{0} \rangle &= \left(1-\frac{1}{768}\frac{\pi^2}{N^2}-\frac{1}{512}\frac{\pi^3}{N^3}-\frac{2419}{9\cdot2^{17}}\frac{\pi^4}{N^4} + \ldots \right) |0 \rangle \nonumber \\
& \qquad + \sqrt{6}\left(\frac{1}{48}\frac{\pi}{N} + \frac{1}{64}\frac{\pi^2}{N^2} + \frac{371}{9\cdot2^{12}}\frac{\pi^3}{N^3} + \ldots\right) |4 \rangle \nonumber \\
& \qquad + 12 \sqrt{70} \left( \frac{1}{96^2}\frac{\pi^2}{N^2} + \ldots\right) |8 \rangle + \ldots,
\end{align}
where we have only written out the terms needed for computing matrix elements to $O(N^{-4})$. Note that we are able to work to \emph{fourth} order despite having only computed the exponent of $U^{\dagger}$ to third order, because the term with the fewest boson operators appearing in the exponent of $U^{\dagger}$ is always $(a^{\dagger 4} - a^4)$: this is a manifestation of the rotational symmetry of the continuum being broken by the square lattice. It can be verified that adding fourth-order corrections to the coefficients in \eqref{eq-U3} do not affect any of the terms in \eqref{eq-perturb-gs}, or any of the results below.

We now seek to compute BZ averages of metric components, using
\begin{equation}
\label{eq-realspace-metric2}
g_{ij} = \text{Re }\tr \left[ P_{\tilde{0}} \hat{r}_i \left(1 - P_{\tilde{0}}\right) \hat{r}_j P_{\tilde{0}} \right ].
\end{equation}
where $P_{\tilde{0}} = | \widetilde{0} \rangle \langle \widetilde{0}|$ projects onto the lowest band of the Hofstadter model, which may be expressed in terms of LL projectors using \eqref{eq-perturb-gs}. Though the form of the operator $U^{\dagger}$ (and hence $| \widetilde{0} \rangle$) was derived in the Landau gauge, the perturbative corrections we incorporate preserve the degeneracy of the Landau levels and therefore are independent of the choice of gauge at our level of approximation. This point is discussed in more detail in Sec.~VI~A of Ref.~\onlinecite{Harper:2014vi}. The calculation of the averaged metric is easiest to carry out in the symmetric gauge, where the quantum-mechanical position operators may be expressed in terms of the LL ladder operators $a, a^{\dagger}$ and angular momentum raising and lowering operators $b^{\dagger},b$ as
\begin{align}
\widehat{x} &= \tfrac{i}{\sqrt{2}}( a - a^{\dagger} -ib - ib^{\dagger}), \\
\widehat{y} &= \tfrac{1}{\sqrt{2}}( a + a^{\dagger} +ib - ib^{\dagger}).
\end{align}
In \eqref{eq-realspace-metric2} these operators appear sandwiched between the ground-state projector and its complement. This simplifies the problem considerably: because $b, b^{\dagger}$ do not mix LLs, no term containing them contributes in \eqref{eq-realspace-metric2}, and we may neglect the angular momentum degree of freedom and work purely in the space of LL indices (as the notation of \eqref{eq-perturb-gs} suggests). Furthermore, because $| \widetilde{0} \rangle$ to $O(N^{-4})$ only has support on LLs $|0\rangle, |4\rangle$ and $|8\rangle$, terms of the form $aa$ and $a^{\dagger}a^{\dagger}$ vanish as well. The only matrix element that contributes to \eqref{eq-realspace-metric2} is
\begin{align}
\label{adag-a-matrix-elm}
\langle \widetilde{0} | a^{\dagger} a | \widetilde{0} \rangle &= \frac{1}{96}\frac{\pi^2}{N^2} +\frac{1}{64}\frac{\pi^3}{N^3} \nonumber \\
& \qquad + \frac{311}{9\cdot2^{11}}\frac{\pi^4}{N^4}+ O\left({N^{-5}}\right).
\end{align}

Eq.~\eqref{adag-a-matrix-elm} suffices to compute the averages of all elements of $g_{ij}$ as $\left<g_{xx}\right> = \left<g_{yy}\right> = 1/2 + \langle \widetilde{0} | a^{\dagger} a | \widetilde{0} \rangle$ and $\left<g_{xy}\right> = 0 + O(1/N^{5})$. Although $g_{xx}$ and $g_{yy}$ do not have the same $\bfk$ dependence, symmetry requires that their BZ averages be the same. 

Because the BZ average of the curvature is always exactly quantized to the Chern number and all higher moments of all quantities (considered as distributions over the BZ) vanish to any order in $1/N$ (due to the fact that our analysis of the WKB wavefunctions in Sec.~\ref{sec-wkb} showed that the amplitude of all $\bfk$ dependence falls off exponentially in $N$), it follows that
\begin{align}
\label{eq-asymp-t}
\langle T \rangle &= 2\left<g_{xx}\right> - 1 \nonumber \\
&\sim  \frac{1}{48}\frac{\pi^2}{N^2} +\frac{1}{32}\frac{\pi^3}{N^3} + \frac{311}{9\cdot2^{10}}\frac{\pi^4}{N^4}+\ldots; \\
\label{eq-asymp-d}
\langle D \rangle &= \left<g_{xx}\right>^{2} - 1/4\nonumber \\
&\sim  \frac{1}{96}\frac{\pi^2}{N^2} +\frac{1}{64}\frac{\pi^3}{N^3} + \frac{313}{9\cdot2^{11}}\frac{\pi^4}{N^4}+\ldots.
\end{align}

\begin{figure}[thb]
\centering
\includegraphics[width=3.0in]{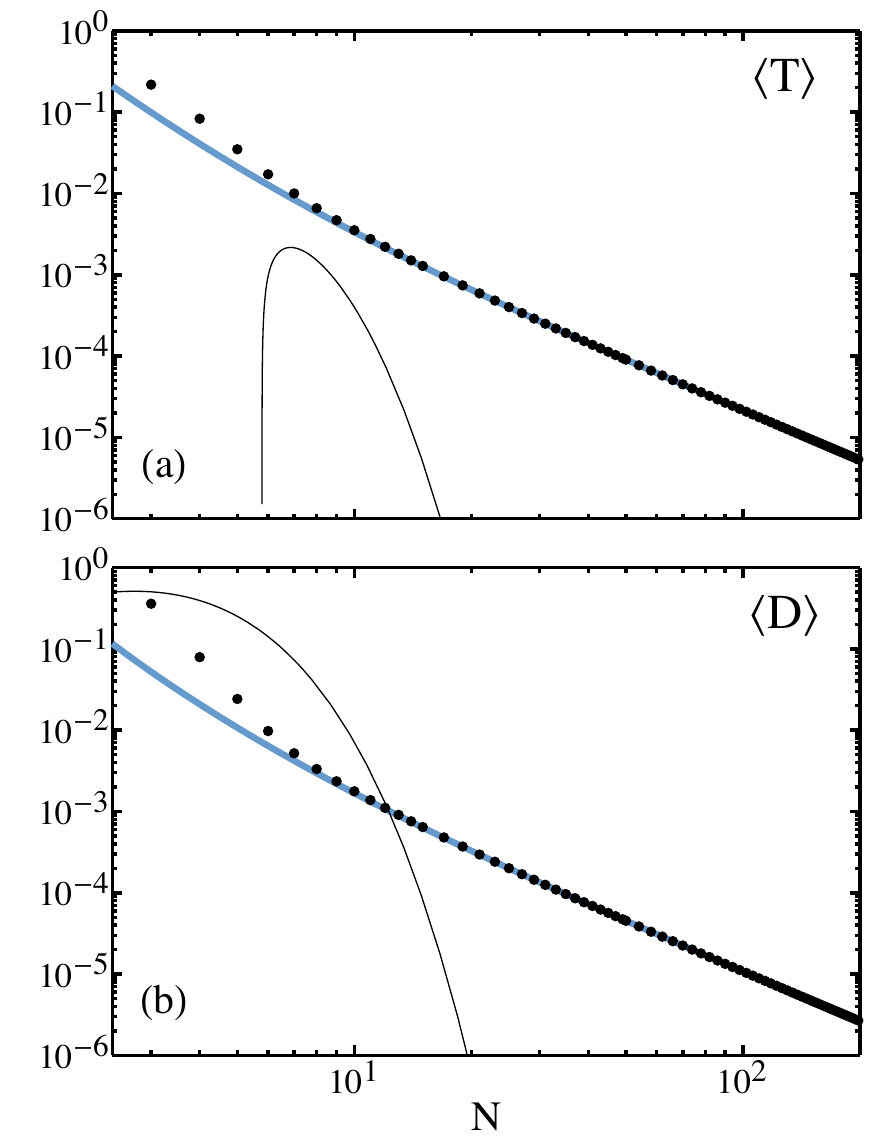}
\caption{\label{fig-trg-detg} (color online). Plot of the BZ-averaged value of the trace \eqref{eq-trineq} and determinant inequalities \eqref{eq-detineq}. Points are obtained from numerically integrated values of $\langle T \rangle$ and $\langle D \rangle$. The blue lines are given by Eqs.~\eqref{eq-asymp-t}, \eqref{eq-asymp-d}. The thin black lines show the fluctuation amplitudes $\tilde{T}$, $\tilde{D}$ plotted in the last row of Fig.~\ref{fig-fluct} for comparison.
}
\end{figure}

Unlike the momentum-space fluctuations of the metric components and Berry curvature, the trace and determinant inequalities have a much slower asymptotic decay which is \emph{polynomial} in $1/N$ (Fig.~\ref{fig-trg-detg}), rather than exponentially small in $N$. Therefore, at large $N$ these conditions should be the dominant factors dictating the stability of FQH-like phases. In practice, this regime occurs for $N$ larger than $\sim 15$, as can be seen in Fig.~\ref{fig-trg-detg}.

\begin{figure}[thb]
\centering
\includegraphics[width=3.0in]{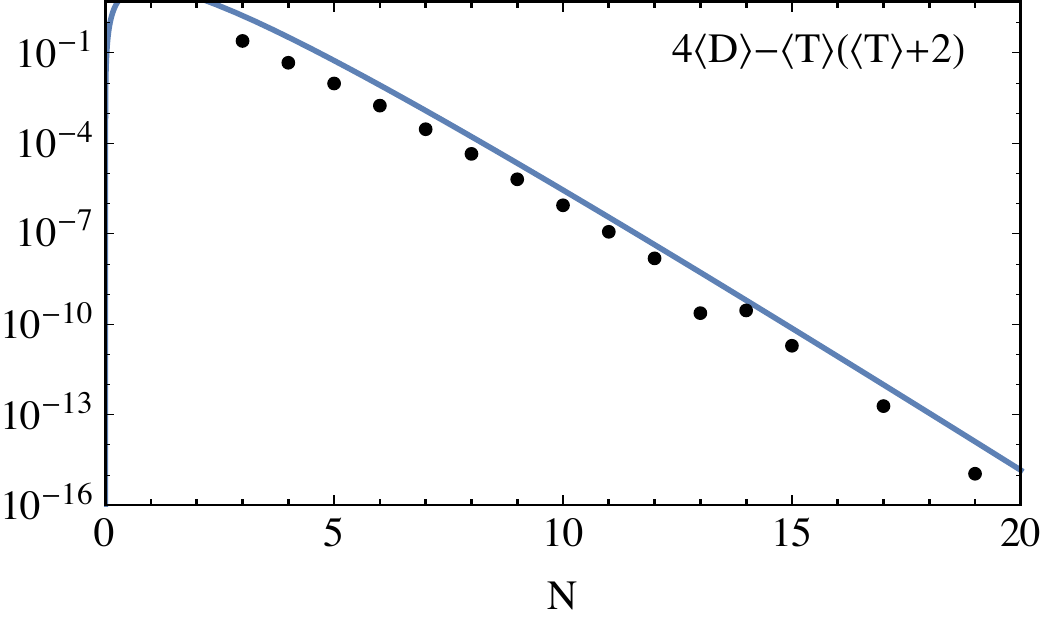}
\caption{\label{fig-parabola} (color online). Plot of the residuals in the implicit relationship between \eqref{eq-asymp-t} and \eqref{eq-asymp-d}. Points are obtained from  values of $\langle T \rangle$ and $\langle D \rangle$ numerically integrated over the BZ. The blue line is a plot of the predicted $N$ dependence, $|A|^{4}N^{4}e^{-2\tilde{\sigma}N}$, as a guide to the eye.}
\end{figure}

Finally, as an overall check on the consistency of our large-$N$ approximation, one may eliminate $\langle g_{xx}\rangle$ from \eqref{eq-asymp-t}, \eqref{eq-asymp-d} to obtain $4\langle D \rangle - \langle T \rangle (\langle T \rangle +2) =0$; the left-hand side of this relation is plotted in Fig.~\ref{fig-parabola}. All neglected terms, namely $\left<g_{xy}\right>$ and BZ fluctuations of terms quadratic in $B$ and $g$, vanish exponentially with the $N$ dependence expected from the WKB calculation of Sec.~\ref{sec-wkb}.

\section{Stability of FQH states}
\label{sec-multiparticle}

\subsection{Scaling of many-body gaps}
\label{sec-scaling}

We investigate the degree to which the above band-geometric criteria influence the stability of an FQH state by adding repulsive interactions and numerically computing the many-body gap for different values of $\phi = 1/N$. Lowering the flux per plaquette changes the relative strength of the interaction, since the size of the system (size of the magnetic unit cell) increases while the number of particles remains constant. In order to compare gaps at different $N$, the strength of the interaction potential must be scaled with $N$ as follows.

In the spirit of the WKB analysis, let $\psi_n, n = 1, \ldots, N$ be a wavefunction defined on the lattice and $\psi(x)$ be its large-$N$ continuum approximation, with $x = n/N$. Under the (nontrivial) assumption that $\psi_n$ has support over almost all tight-binding orbitals, the requirement that both $\psi_n$ and $\psi(x)$ be normalized to unity dictates the scaling $\psi_n \sim \sqrt{N}\psi(x)$. Consideration of the matrix elements of a two-body $\delta$-function interaction $V(x) = V_{0}\delta(x)$ shows that
\begin{align*}
\left<\psi^3,\psi^4\right|\hat{V}\left|\psi^1,\psi^2\right> &= V_0 \int dx\, \psi^{3}(x)^{\ast} \psi^{4} (x)^{\ast}\psi^1(x)\psi^2(x)\\
 &\sim N V_0 \sum_{n} \psi^{3\ast}_n \psi^{4\ast}_n \psi^1_n \psi^2_n,
\end{align*}
so that the strength of the corresponding discrete on-site interaction should scale as $V_{\text{disc}} \sim V_{0} N$. Similar considerations show that for the two-body nearest-neighbor repulsion used to stabilize the fermionic Laughlin state and for the three-body delta-function interaction stabilizing the Moore-Read state, the leading scaling should be $V_{\text{disc}} \sim V_{0} N^{2}$. Because the single-particle bandwidth vanishes exponentially fast with $N$, the many-body gap should be scaled in the same way, and we define $\Delta_{\text{sc}} = N\Delta$ for the bosonic Laughlin state and $\Delta_{\text{sc}} = N^{2}\Delta$ for the fermionic Laughlin and Moore-Read states, where $\Delta$ is the gap obtained with a fixed interaction strength. Similar scaling arguments can be used to compare FCI models with different numbers of bands \cite{Jackson:tsinVXaa}.

\subsection{Many-body computations}
\label{sec-manybody}

In order to study the influence of band geometry on FCI phases, we carried out exact numerical diagonalization of a repulsive interaction Hamiltonian projected to the lowest Hofstadter band for $N_p = 8$ particles partially filling the lowest band. To study the bosonic Laughlin state at $\nu = 1/2$, we used a lattice of $4\times4$ unit cells and a two-body on-site repulsion; for the fermionic Laughlin state at $\nu = 1/3$ we used a $4 \times 6$ lattice with a two-body nearest-neighbor repulsion, and for the bosonic Moore-Read state at $\nu =1$ we used a $2 \times 4$ lattice with a three-body on-site repulsion. For geometric reasons, we used system sizes of the form $N=m^{2}$ up to $N=169$ for the bosonic Laughlin state, $N=6m^{2}$ up to $N=294$ for the fermionic Laughlin state, and $N=2m^{2}$ up to $N=162$ for the bosonic Moore-Read state. 

For each system, we verified that the many-body ground state had nontrivial topological order corresponding to the appropriate FQH state via properties of energy and entanglement spectra. We required that the many-body spectrum exhibit a quasidegenerate ground state, with the gap $\Delta$ to excited states larger than the spread in ground-state energies. We compute the particle entanglement spectrum by tracing out four particles from the density matrix formed by an equal superposition of all ground states, and required that this spectrum be gapped, with the counting of eigenvalues below the gap in each momentum sector given by the appropriate counting rules \cite{Regnault:2011bu,Bernevig:2012ka} for the corresponding FQH state.

\begin{figure}[bth]
\centering
\includegraphics[width=3.1in]{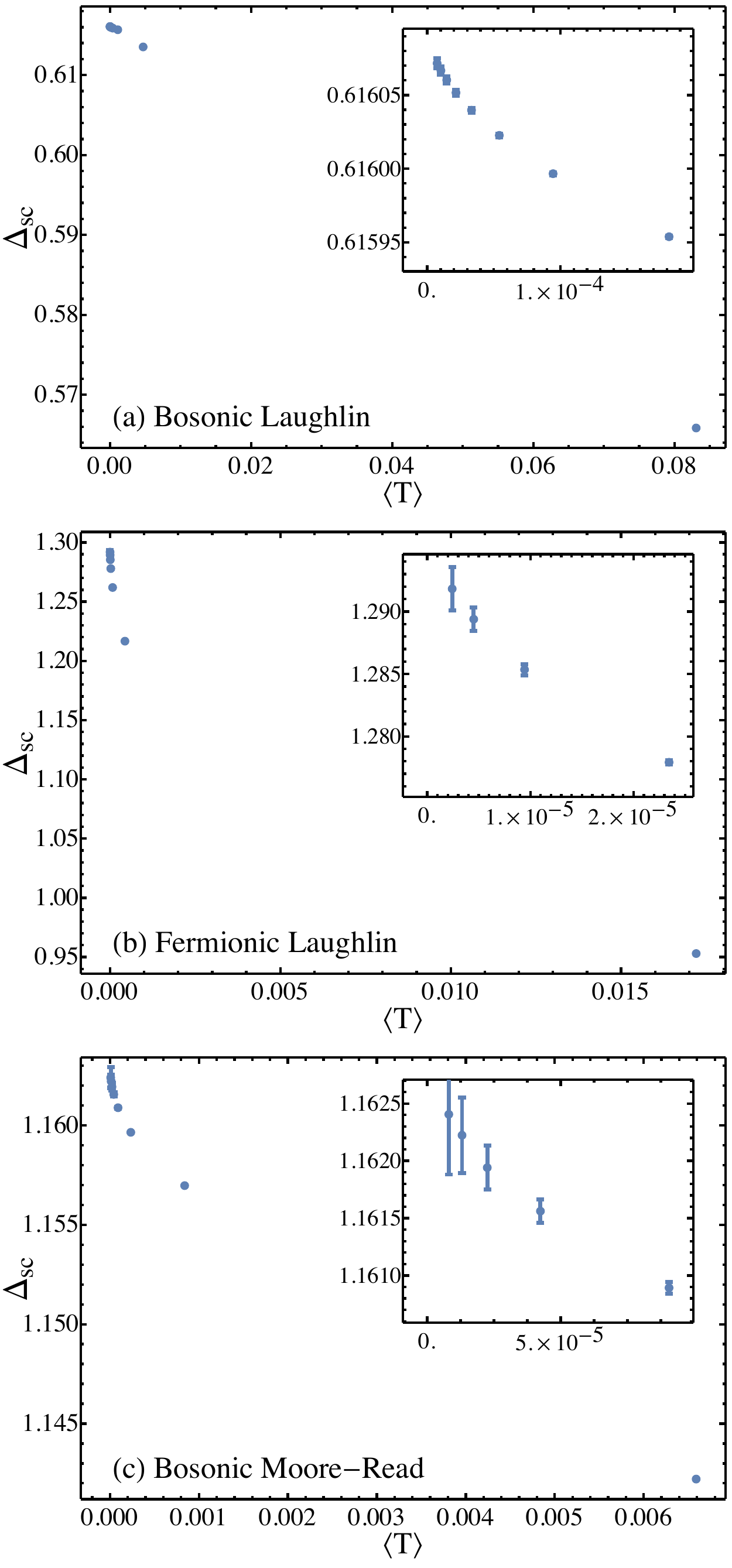}
\caption{\label{trg-gaps} (color online). Scaled many-body gaps as a function of BZ-averaged trace inequality \eqref{eq-trineq} for FQH-like states at various values of $N$. Insets show behavior of points at larger values of $N$. Error bars reflect the finite accuracy of our diagonalization code. (a). Laughlin state of $N_{p}=8$ bosons at $\nu = 1/2$. (b). Laughlin state of $N_{p}=8$ fermions at $\nu = 1/3$. (c). Moore-Read state of $N_{p}=8$ bosons at $\nu = 1$.}
\end{figure}

In Fig.~\ref{trg-gaps} we plot the scaled gaps as a function of $\left<T\right>$, the Brillouin zone average of the trace inequality \eqref{eq-trineq}. We note a clear correlation between an increasing value of the gap and increasing saturation of the inequality, with the gap continuing to increase even for values of $N$ for which BZ fluctuations of $B$ and $g$ are negligible (as shown in Fig.~\ref{fig-fluct}). Apart from a change in horizontal scale, dependence of the scaled gaps on $\langle D \rangle$ is visually identical: comparing \eqref{eq-asymp-t} with \eqref{eq-asymp-d} shows that, in the large-$N$ limit, $\langle D \rangle \approx 2 \langle T \rangle$ up to terms of $O(N^{-4})$.

\section{Discussion}
\label{sec-discussion}

The Hofstadter model provides an ideal laboratory for studying FCI phenomena: the existence of a controlled limit in which its spectrum converges to continuum LLs, which makes the model amenable to perturbative expansion in $1/N$ and allows a controlled study of the relationship between band geometry and FQHE physics. In this paper, we have shown that there is a natural distinction between effects that are nonperturbatively small in $1/N$, such as the BZ fluctuations of the Berry curvature and quantum metric, and effects which have a perturbative expansion in $1/N$, such as the trace and determinant conditions. The behavior of the many-body gaps obtained via exact diagonalization indicates that the latter effects dominate for large $N$. In the future, it would be interesting to study models where the role of the determinant inequality can be isolated. 

The Hofstadter model has been realized experimentally in optical lattices of cold atoms \cite{Aidelsburger:2013ew,Miyake:2013jw,Aidelsburger:2014hm} and graphene superlattices \cite{Dean:2013bv,Ponomarenko:2013hl}. The geometrical criteria for the stability of Chern bands are readily computable single-particle properties that act as a meaningful proxy for the many-body gap, and hence may serve as a useful guide in selecting experimental parameters and couplings which are most favorable to the existence of a stable FQH state. In this sense, the present work complements Ref.~\onlinecite{Jackson:tsinVXaa}, in which the effects of band geometry on the many-body gap were measured by varying the couplings of several FCI models. In that study, BZ fluctuations of the Berry curvature were found to have the largest influence on the gap, with the trace condition a subdominant factor.

In this paper we have presented a perturbative treatment of the band geometry of the Hofstadter model at $\phi = 1/N$ in the limit $N \to \infty$, but our analysis may be extended to the most general case considered in \cite{Harper:2014vi}, namely $\phi = P/Q + M/N$ with $N \gg M$. Our own single- and many-particle computations suggest that the case $\phi = M/N$ is qualitatively similar to the results discussed above, with all quantities taking values which interpolate among those obtained at $M=1$. Different physics is encountered when perturbing around nonzero $P/Q$, since the lowest band now has Chern number $c_{1}=Q >1$. FCI states in bands with $c_{1}>1$ may be mapped onto multilayer FQH states \cite{Barkeshli:2012kw}, albeit with significant distinctions \cite{Liu:2012ek,Sterdyniak:2013du,Wu:2013ii}. The specific case of the Hofstadter model (and generalizations) at $\phi = P/Q$ has been examined in \cite{Wang:2013he,Scaffidi:2014tg,Wu:2015id,gunnar-unpublished}. 

One could also extend our analysis to excited bands of the Hofstadter model, corresponding to higher LLs. Finally, obtaining an analytic formula for the quantitative dependence of the gap (or other many-body observables) on band-geometric parameters remains an important open problem.

\begin{acknowledgments}
We thank G.~M\"oller for helpful discussions and for sharing Ref.~\onlinecite{gunnar-unpublished} in advance of publication. We also thank F.~Harper for sharing his unpublished notes with us and correcting mistakes in earlier drafts of this paper. R.~R. thanks S.~H.~Simon and F.~Harper for useful discussions and a fruitful collaboration. R.~R. and D.~B. acknowledge support from the NSF under CAREER DMR-1455368 and the Alfred P. Sloan foundation.
\end{acknowledgments}

%%%%%%%%%%%%%%%%%%%%%%%%%%%%%%%%%%%%%%%%%%%%%%%%%%
\bibliographystyle{apsrev4-1}
%\bibliography{hofstadter,hofstadter-manual}
% paste .bbl


%%%%%%%%%%%%%%%%%%%%%%%%%%%%%%%%%%%%%%%%%%%%%%%%%%
\appendix

\section{Closed-form band geometry for $N=3$}
\label{app-n3}

In \cite{Barnett:2012id}, an analytic expression was obtained for the Berry curvature of an arbitrary three-band Hamiltonian. In this section, we obtain analogous expressions for components of the quantum metric, for the purpose of determining what statements may be made about the full momentum dependence of the quantum metric of the Hofstadter model at general $N$.

We follow the notational conventions of Ref.~\onlinecite{Barnett:2012id}. Just as an arbitrary two-band Hamiltonian may be written in terms of Pauli matrices and its eigenstates described using a Bloch spin, we may write any three-band Hamiltonian as  $H(\bfk) = a(\bfk) + \bfb(\bfk) \cdot \boldsymbol{\lambda}$, with the projector onto the $\alpha$th band given by $P_{\alpha}=\frac{1}{3}(1+ \sqrt{3} \bfn_{\alpha}(\bfk) \cdot \boldsymbol{\lambda})$, for $\alpha = 1, 2, 3$. Here $ \boldsymbol{\lambda}$ is an eight-component vector, the entries of which are the Gell-Mann matrices forming the fundamental representation of $SU(3)$. These obey 
\begin{equation}
\lambda_{a}\lambda_{b}= \tfrac{2}{3} \delta_{ab} + (d_{abc} + i f_{abc})\lambda_{c}.
\end{equation}
Using these structure constants we may define antisymmetric and symmetric vector products on eight-component vectors $\bfx, \bfy$ as $(\bfx \times \bfy)_{a} = f_{abc}\bfx_{b}\bfy_{c}$ and $(\bfx \ast \bfy)_{a} = \sqrt{3}d_{abc}\bfx_{b}\bfy_{c}$, respectively.

Barnett, Boyd and Galitski parameterize the relationship between $H$ and $P_{\alpha}$ via $\mathbf{n}_{\alpha} = \xi_{\alpha}(\gamma_{\alpha}\mathbf{b} + \mathbf{b} * \mathbf{b})$, with 
\begin{equation*}
\xi_{\alpha} = \frac{1}{\gamma_{\alpha}^2 - |\mathbf{b}|^2}; \qquad \gamma_{\alpha} = 2|\mathbf{b}|\cos \theta_{\alpha},
\end{equation*}
where
\begin{equation}
\label{eq-barnett-theta}
\theta_{\alpha} = \frac{1}{3}\cos^{-1}\left(\frac{\mathbf{b}\cdot\mathbf{b}*\mathbf{b}}{|\mathbf{b}|^3}\right) + \frac{2 \pi}{3} \alpha,
\end{equation}
and, for clarity, we've suppressed $\bfk$ dependence of all quantities. The fact that $P_{\alpha}$ is a projector implies $|\bfn_{\alpha}|^{2}=1$ and $\bfn_{\alpha} \ast \bfn_{\alpha} = \bfn_{\alpha}$; the fact that it projects onto an eigenstate of $H$ means $[P_\alpha, H]=0$, which implies $\bfb \times \bfn_{\alpha} =0$.

The $\bfb$ vector corresponding to the $N=3$ Hofstadter Hamiltonian in the Landau gauge is given (up to an irrelevant permutation of rows and columns) following Eq.~(4) of Ref.~\onlinecite{Barnett:2012id}:
\begin{align*}
\bfb(\bfk) &= \left\{ \cos k_{x}, \sin k_{x}, \tfrac{\sqrt{3}}{2}\sin k_{y}-\tfrac{3}{2}\cos k_{y}, \cos k_{x}, \right. \\
& \qquad \left. -\sin k_{x}, \cos k_{x}, \sin k_{x}, \tfrac{\sqrt{3}}{2}\sin k_{y}+\tfrac{3}{2}\cos k_{y} \right\},
\end{align*}
which is of constant length $|\bfb|^{2}=6,$ implying $\bfb \cdot d\bfb =0$. Additionally,
\begin{equation}
\bfb \cdot \bfb \ast \bfb = 3^{3/2}[\cos (3k_{x}) + \cos (3k_{y})],
\end{equation}
therefore, $\theta_{\alpha}$ and the other coefficients only depend on momentum through the function $\cos (3k_{x}) + \cos (3 k_{y})$. This is the $\cos (Nk_{x}) + \cos (Nk_{y})$ momentum dependence found for the Berry curvature in Ref.~\onlinecite{Harper:2014vi}.

Components of the quantum metric may be computed as
\begin{align}
g_{ij}^{\alpha} &= \frac{1}{2}\tr \left(P_{\alpha}\partial_{k_i}P_{\alpha}\partial_{k_j}P_{\alpha}\right)\\
&= \frac{1}{3}\partial_{k_i}\bfn_{\alpha}\cdot\partial_{k_j}\bfn_{\alpha},
\end{align}
where we take $\alpha = 1$ for the lowest band. The fact that this expression involves the dot product instead of the antisymmetric product means that convenient cancellations arising in the computation of the Berry curvature do not happen here. We find 
\begin{align}
\label{eq-analytic-gxx}
g_{xx} &= \frac{1}{12[2\cos (2\theta) + 1 ]^{4}}  \\
& \quad \times \left\{ 5+16\cos^{2} \theta \left[ 1 + 5 \cos (2\theta) + \cos (4\theta)\right]\right. \nonumber \\
& \quad \qquad -8 \sqrt{2} \cos \theta \left[1+ 2\cos (2\theta) \right]^{2} \cos (3k_{x}) \nonumber \\
&\quad \qquad  \left. +\left[ 5+4 \cos (2 \theta) \right]\cos (6k_{x}) \right\}, \nonumber
\end{align}
with $\theta \equiv \theta_{1}(\bfk)$ given by \eqref{eq-barnett-theta}. The expression for $g_{yy}$ is analogous, with the explicit factors of $k_{x}$ appearing in \eqref{eq-analytic-gxx} replaced by $k_{y}$, so that
\begin{align}
\tr g &= \frac{2}{3[2\cos (2\theta) + 1 ]^{4}} \\
& \quad \times \left\{ 6 +4 \cos (2\theta) - \cos (4\theta)\right. \nonumber \\
& \quad \qquad \left.-\left[4 \cos (2 \theta) +5\right] \cos (3k_{x})\cos (3k_{y}) \right\}.  \nonumber
\end{align}
The momentum dependence of $\tr g$ involves both the sum $\cos (Nk_{x}) + \cos (Nk_{y})$ (through the $\bfk$-dependent variable $\theta$) and the product $\cos (Nk_{x}) \cos (Nk_{y})$. We now show that the same situation holds for $\det g$: using
\begin{equation}
g_{xy}= -\frac{[4 \cos (2 \theta) + 5]\sin (3k_{x}) \sin (3k_{y})}{6[2\cos (2\theta) + 1 ]^{4}},
\end{equation}
we have
\begin{align}
\det g &= \frac{1}{36[2\cos (2\theta) + 1 ]^{6}}  \\
& \quad \times \left\{ 29 -12 \cos (2\theta) - 48\cos (4\theta)\right. \nonumber \\
& \quad \qquad - 12\cos (6\theta)- 2\cos (8\theta) \nonumber \\
& \quad \qquad \left.+6\left[2\cos (2 \theta) +1\right] \cos (3k_{x})\cos (3k_{y}) \right\}. \nonumber
\end{align}
As proved in Sec.~\ref{app-symmetry}, this is the most general form of $\bfk$ dependence a function with the lattice symmetries of $\tr g$ and $\det g$ is allowed to have. Despite being fully consistent with that proof, this result was somewhat unexpected: the Berry curvature has the same lattice symmetries, is defined as a similar combination of $\bfk$ derivatives of Bloch functions, and enters into the band geometry on an equal footing as the quantum metric \cite{Roy:2012vo}, yet only depends only on the sum $\cos (Nk_{x}) + \cos (Nk_{y})$. 

\begin{figure}[hb]
\centering
\includegraphics[width=3.25in]{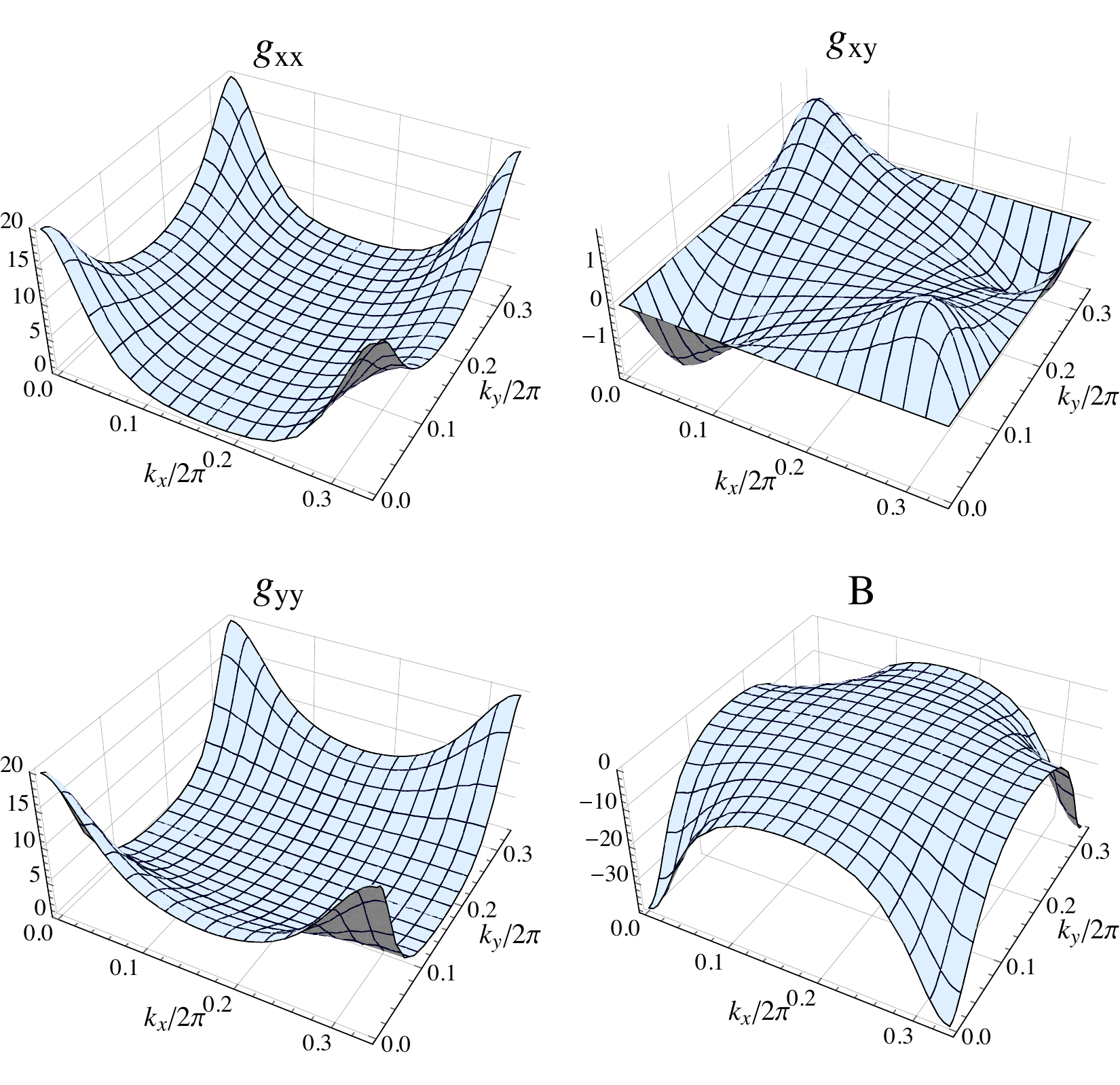}
\caption{\label{fig-metric3d} (color online). Plots of the metric components and Berry curvature over a minimal periodic cell of the BZ (of size $2\pi/N \times 2 \pi/N$) for the Hofstadter model at $N=3$.}
\end{figure}

In Fig.~\ref{fig-metric3d} we plot the curvature and each component of the metric for $\phi=1/3$, where the fluctuations are largest in magnitude. The lattice symmetry properties proved above are evident.

%%%%%%%%%%%%%%%%%%%%%%%%%%%%%%%%%%%%%%%%%%%%%%%%%%

\end{document}